# Isotropic to nematic transition in alcohol ferrofluids of barium hexaferrite nanoplatelets

Patricija Hribar Boštjančič[1,2*], Žiga Gregorin[1], Nerea Sebastián[1], Natan Osterman[1,3], Darja Lisjak[4], Alenka Mertelj[1*]

[1] Department of Complex Matter, Jožef Stefan Institute, Jamova cesta 39, SI-1000 Ljubljana, Slovenia

[2] Jožef Stefan International Postgraduate School, Jamova cesta 39, SI-1000 Ljubljana, Slovenia

[3] Faculty of Mathematics and Physics, University of Ljubljana, Jadranska ulica 19, SI-1000 Ljubljana, Slovenia

[4] Department for Materials Synthesis, Jožef Stefan Institute, Jamova cesta 39, SI-1000 Ljubljana, Slovenia

* Corresponding author: Patricija Hribar Boštjančič, Alenka Mertelj

Tel.: +386 1 477-3937

E-mail address: patricija.hribar.bostjancic@ijs.si and alenka.mertelj@ijs.si

**Abbreviations**
BHF: Barium hexaferrite; NPLs: Nanoplatelets; POM: Polarizing optical microscopy; DBSA: Dodecylbenzenesulfonic acid; $c_{dis}$: Molar concentration of dissolved DBSA; TEM: Transmission electron microscopy; MO: Magneto-optic.

**Abstract**

Alcohol ferrofluids made of ferrimagnetic barium hexaferrite (BHF) nanoplatelets (NPLs) form a unique example of a dipolar fluid – a liquid magnet. Its formation is induced at a high enough concentration of the NPLs. The key interactions between the NPLs are long-ranged dipolar magnetic and screened anisotropic electrostatic. Herein, we report the results on tuning the isotropic-nematic phase transition (i.e., the NPLs threshold concentration) in 1-butanol ferrofluids of BHF NPLs by affecting the interactions in the ferrofluids. The threshold concentration was determined by polarizing optical microscopy (POM) combined with a system for magnetic field manipulation and was in the range between 4.6 and 6.6 % (v/v) depending on the ferrofluid. We observed that the threshold concentration decreased for 0.6 % (v/v) with a larger mean diameter of the NPLs, up to 0.6 % (v/v) with increased ionic strength of the ferrofluid, and for ~ 2 % (v/v) with higher saturation magnetization of the NPLs. We showed that by tuning the parameters affecting the ferrofluid's interplatelet interactions, we can alter the threshold concentration for the liquid magnet formation. The results elucidate the importance of a delicate balance between the repulsive screened electrostatic and attractive dipolar magnetic interactions for the liquid magnet formation.

# 1. Introduction

Liquid-crystal behavior in colloid suspensions of anisotropic particles has been observed already at the beginning of the 20$^{th}$ century [1–3]. Langmuir studied a plate-like clay particle suspension (the term suspension will be reserved for suspensions of nonmagnetic particles), which separated in two distinct phases at the equilibrium, i.e., isotropic and nematic phase [1]. Since then, several systems made of the plate- [4–9] and rod-like [10,11] particles were studied. Orientational correlations between anisotropic particles have a significant impact on the phase behavior and the nematic phase formation in colloidal suspensions [12]. The particle charge and ionic strength of the suspension have been shown as essential parameters for tuning the effective aspect ratio, i.e., the ratio between the thickness and width of the particles [8].

Shuai et al. recently reported the first experimental observation of nematic ordering of ferrimagnetic BHF NPLs in 1-butanol [13]. The reason that enabled such ordering is the anisotropy of the NPLs and the magnetic moment perpendicular to the basal NPL's plane. With the nematic ordering of the NPLs, the magnetic moments order simultaneously, and in such a way, a macroscopic ferromagnetic fluid – a liquid magnet is formed. The threshold concentration for the ferromagnetic nematic ordering was determined where it is expected due to platelet shape (28 % (v/v)) [13]. However, a later study indicates that the strong ferromagnetic correlations between the magnetic NPLs can be responsible for the formation of the ferromagnetic phase at much lower volume concentrations [14]. The ferromagnetic ordering of the BHF NPLs does not depend only on the platelet shape and strength of magnetic interaction but also on the anisotropic electrostatic interaction energy, needed for the colloidal stability [14]. The colloidal stability is crucial for the preparation of highly concentrated ferrofluids with the ordered NPLs. The surfactant dodecylbenzenesulfonic acid (DBSA) ensures colloidal stability by forming a double-layer at the surface of the NPLs (Fig. 1(a) and (b)) [15], where the second layer is positively charged in 1-butanol and defines the surface charge of the NPLs. Some fraction of DBSA (~ 5 mM) is dissolved ($c_{dis}$) in the ferrofluid and contributes to the ionic strength, determining the Debye screening length [16]. The understanding how the parameters (i.e., anisotropic shape, magnetic dipole moment, and anisotropic electrostatic interaction) contribute to the formation of the ferromagnetic phase is still missing.

Here, we report a study on the phase behavior of the ferrofluids of BHF NPLs in 1-butanol. These ferrofluids present a system with complex interactions. We studied the phase behavior by changing the ionic strength of the ferrofluid (DBSA concentration) and magnetic dipole interaction strength (NPLs' magnetization and size). The choice of the materials used for the study was limited to the platelets with the sizes and magnetization from which stable suspensions can be prepared.

## 2. Experimental

### 2.1 Materials

Barium nitrate (99.95 %), iron(III) nitrate nonahydrate (98+ %), scandium(III) nitrate hydrate (99.9 %), indium(III) nitrate hydrate (99.99 %), sodium hydroxide (98 %), and DBSA (97 %) were obtained from Alfa Aesar. Nitric acid (65 %) was purchased from Sigma-Aldrich and 1-butanol (99.4 %) from Baker Analyzed ACS, J.T. Baker.

### 2.2 Synthesis and preparation of the ferrofluids

The In- or Sc-substituted BHF NPLs were synthesized hydrothermally at 245 °C using a well-established procedure [17]. In- and Sc-substituted BHF NPLs were prepared to tune the saturation magnetization and obtain NPLs with comparable size distribution and different magnetization. For the preparation of the BHF NPLs, the barium, iron(III) and scandium(III) (or indium(III)) nitrates were dissolved in deionized water in a molar ratio of 1:4.5:0.5 and precipitated from the solution after the addition of excess sodium hydroxide. After precipitation, DBSA was added to the mixture (to obtain a concentration of 23 mM) that was subsequently transferred to an autoclave (Parr Instruments). The autoclave was heated up to 245 °C with a heating rate of 3 °C/min. When the temperature was reached, the autoclave cooled down to room temperature naturally. The NPLs were centrifuged and washed with water. The surface of the NPLs was modified with DBSA following the procedure described in [15]. Briefly, the NPLs were dispersed in water, the pH was adjusted to 1.5 with 5 M nitric acid, and the mixture was stirred at boiling point for 2 h. The NPLs were washed with water and acetone and dried to remove acetone residues. The dried NPLs were then dispersed in 1-butanol with an ultrasonic probe (Sonics Vibra-cell) at 100 W and pulse 1 s on, 1 s off to prepare a ferrofluid with a concentration of approximately 3 % (v/v). The as-prepared ferrofluid was concentrated with centrifugation at 20 817 rcf for 15 minutes. The centrifuged ferrofluid formed a concentration gradient and was divided into three differently concentrated phases (Fig. 1(c)) with different size distribution. The ferrofluids were centrifuged several times until the middle, and the bottom part formed the nematic phase, while the upper part remained isotropic (for the Sc-substituted NPLs). In-substituted NPLs ordered into nematic phase only in the bottom part of the ferrofluid, leaving the middle and the top part in the isotropic phase.

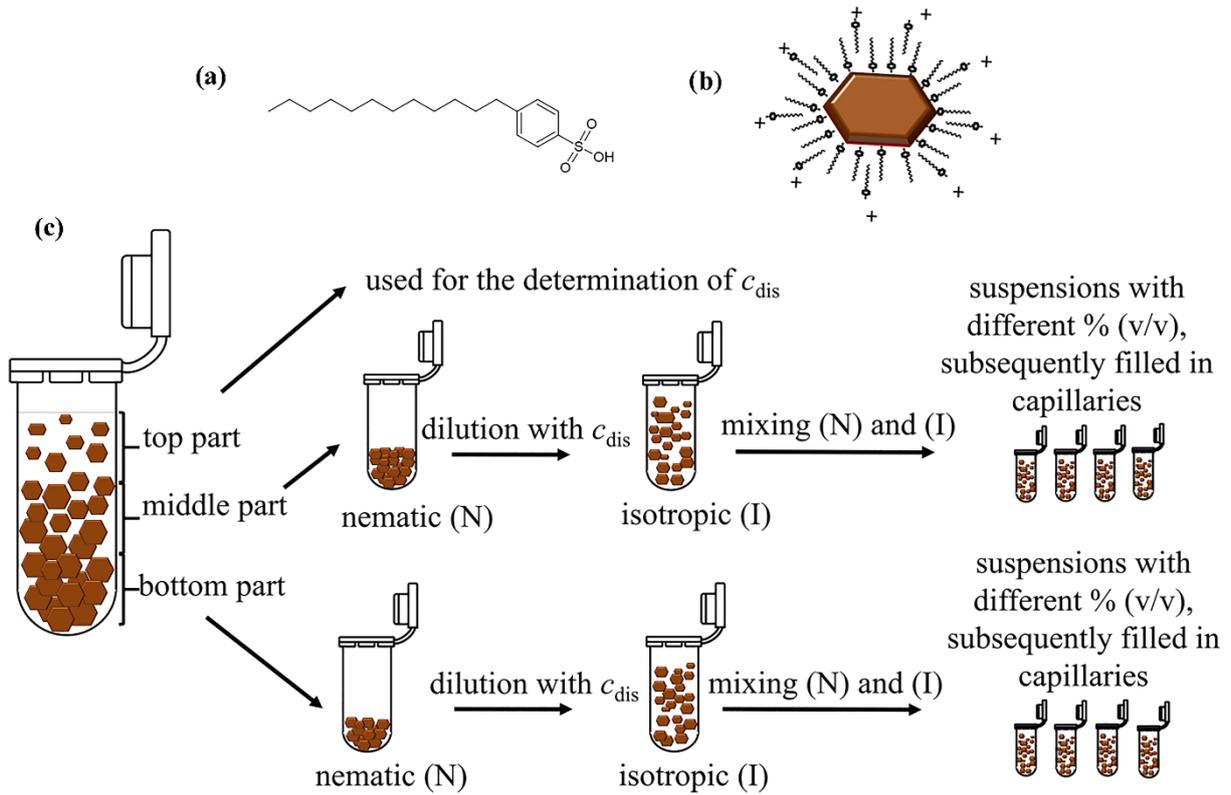

**Fig. 1.** (a) Molecular structure of DBSA, (b) schematic presentation of the DBSA double-layer around an NPL, and (c) scheme of the sample preparation after centrifugation – division in three parts.

The nematic ferrofluids were collected, and the threshold concentration was determined. For this purpose, the nematic ferrofluids were diluted back to the isotropic phase with a 1-butanol solution of DBSA with a concentration $c_{dis}$ (4.5 mM for Sc- and 5 mM for In-sample). The procedure for the determination of $c_{dis}$ is described in Section 2.3. The dilution with a DBSA solution ($c_{dis}$) allowed to obtain sequences of ferrofluids with varying concentrations of NPLs while maintaining unchanged electrostatic interaction energy. The as-prepared isotropic ferrofluid was mixed with the leftover of the as-separated nematic ferrofluid in different ratios to reach the threshold concentration (Fig. 1(c)). Note that the comparison was made between the same parts (i.e., middle to middle, bottom to bottom).

## 2.3 Characterization of materials

The size and shape of the NPLs were determined from transmission electron microscopy (TEM) (Jeol JEM-2100) images (Fig. 2). The TEM samples were prepared by drop-deposition of diluted ferrofluid on a Cu-grid-supported perforated carbon foil. At least 350 particles were measured from the TEM images using a DigitalMicrograph Gatan Inc. software for the particle size distribution. The size distributions (of equivalent diameters, i.e., a disk diameter with the same surface) are depicted in Fig. 3, and the average equivalent diameters are listed in Table 1. The term size will be in the further text used for equivalent diameter.

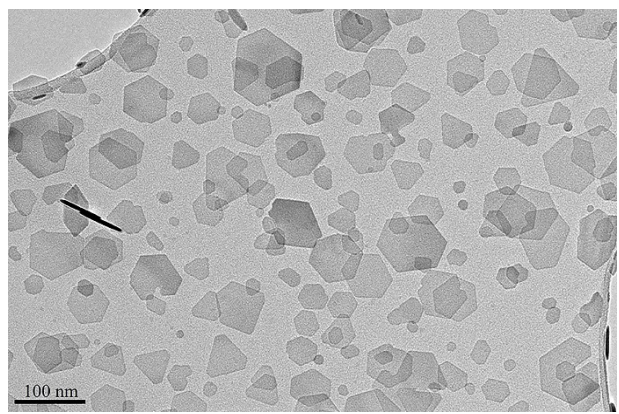

**Fig. 2.** TEM image of the BHF NPLs.

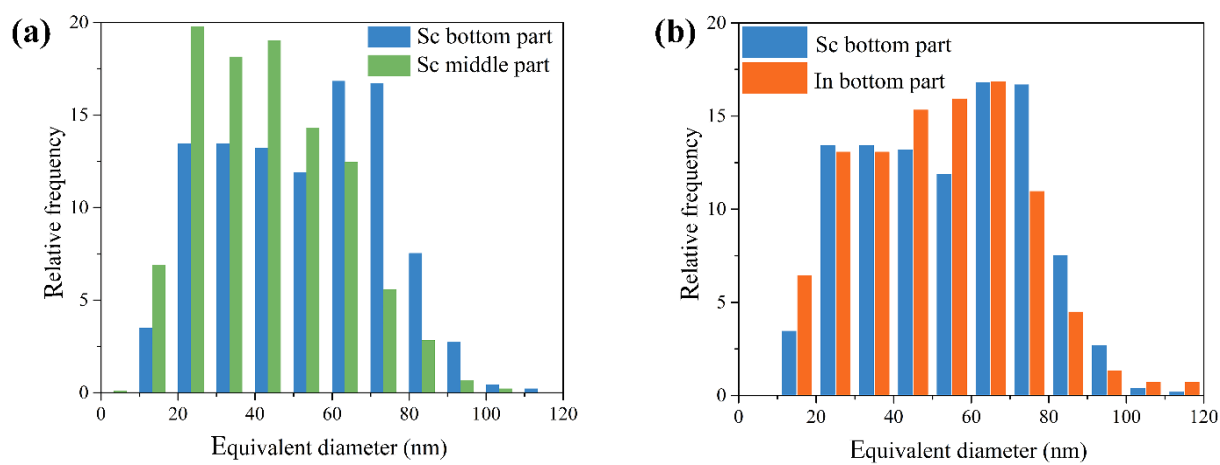

**Fig. 3.** Size distributions of the NPLs (a) comparison of middle and bottom parts of Sc-substituted sample, (b) comparison of bottom parts of Sc- and In-substituted sample.

**Table 1.** Mean equivalent diameter, saturation magnetization, and initial volume concentration for the nematic ferrofluids of the Sc- and In-substituted NPLs.

|  | Sc-substituted NPLs | | In-substituted NPLs |
| --- | --- | --- | --- |
| Part of the centrifuged ferrofluid | bottom | middle | bottom |
| Mean equivalent diameter (nm) | 54 ± 21 | 45 ± 18 | 52 ± 22 |
| Saturation magnetization (Am$^2$/kg) | 38 | 39 | 29 |
| Initial volume concentration (%) | 7.4 | 5.2 | 8.2 |

Magnetic properties of dried NPLs were measured with a vibrating-sample magnetometer Lakeshore 7404. The concentration of $c_{dis}$ in the ferrofluids was determined from the conductivity (Conductometer Knick−Portamess) of the top part (isotropic) after the centrifugation. A standard addition method was applied since the conductivity of the ferrofluid is linearly dependent on the $c_{dis}$ and can be described as $\sigma = a + (b \times (c_{dis}))$, where $\sigma$ is conductivity and $c_{dis}$ can be calculated accordingly. The DBSA solutions with $c_{dis}$ were used for the dilution of nematic ferrofluids (see Section 2.2).

Determination of the ordering in the ferrofluids was done using POM, which is a standard technique for the study of birefringent materials. It allowed to directly discern between isotropic and birefringent nematic phases (see Supplementary Information S.1). External sources did not promote ordering due to compensation of the static environmental magnetic field with a custom-made 3-axis magnetic field manipulation system to field strengths below 1 μT. The ferrofluids with different concentrations were transferred to rectangular capillaries (0.05 mm × 1.00 mm, length 50 mm), sealed and observed by POM with crossed polarizers at 45° to the long axis of the capillary. Additionally, the birefringence of the different samples, directly related to the degree of the nematic ordering, was calculated (see Supplementary Information S.1) to determine the threshold concentration. The results are presented in the graph in Fig. 11.

## 3. Results

**3.1 Effect of the NPLs size**

Our centrifugation procedure (section 2.2) allowed us to separate the nematic ferrofluids of the Sc-substituted NPLs (middle and bottom part) by their concentration. The bottom contained higher concentration of the NPLs (7.4 % (v/v)) with larger mean equivalent diameter (($54 \pm 21$) nm) than the middle (5.2 % (v/v), ($45 \pm 18$) nm) (see Table 1). The Sc-middle turned out to be right at the threshold concentration. Even the smallest dilution resulted in the transition to the isotropic phase. The Sc-bottom was, on the other hand, high above the threshold concentration (Fig. 4(a)). We estimated with dilution that the Sc-bottom's threshold concentration was ($4.6 \pm 0.1$) % (v/v). Close to the threshold concentration, i.e., 4.3 % (v/v), we observed the isotropic-nematic phase coexistence, where the nematic phase formed string-like structures embedded in the surrounding isotropic phase (Fig. 4(b)).

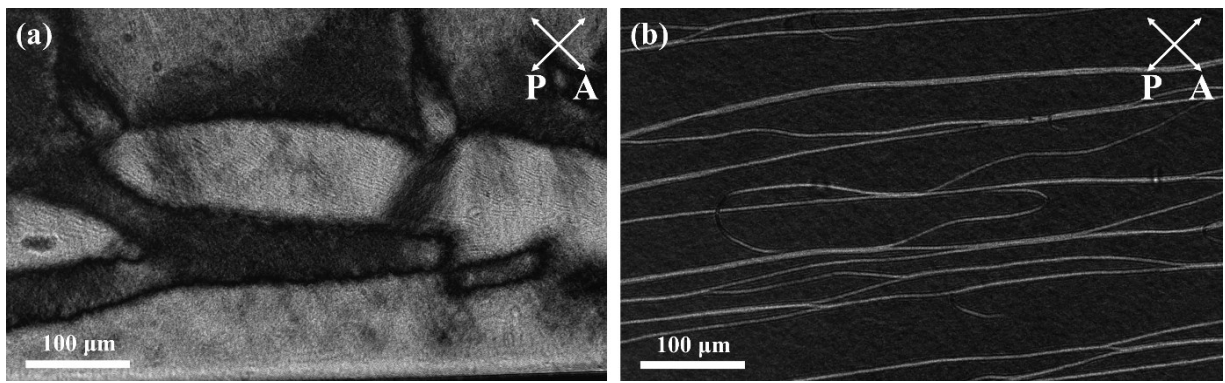

**Fig. 4.** Sc-bottom part: (a) magnetic domains in the nematic sample at 4.7 % (v/v), (b) isotropic-nematic phase coexistence at the concentration of 4.3 % (v/v).

## 3.2 Effect of the DBSA concentration in the ferrofluids

The effect of the ionic strength on the isotropic-nematic threshold concentration was evaluated using the Sc-middle and Sc-bottom with the initial threshold concentration of 5.2 % (v/v) and (4.6 % ± 0.1) % (v/v), respectively. We subsequently increased the $c_{dis}$, while the % (v/v) of the ferrofluid decreased due to the DBSA solution (1.55 M) additions. Higher $c_{dis}$, i.e., > 28 mM decreased the threshold concentration of the original Sc-middle for ~ 0.2 % (v/v) (see Fig. 5 and Fig. 6). The same behavior was observed for the Sc-bottom. The threshold concentration decreased for ~ 0.6 % (v/v) at $c_{dis}$ = 37 mM. Thus, an increase of the $c_{dis}$ decreases the threshold concentration. The $c_{dis}$ affects the Debye length $\lambda_D$ and, consequently, the range of the electrostatic interaction that, in our case, also affects the strength of magnetic interaction between the NPLs. The closer the distance between the neighboring NPLs, the stronger the magnetic interaction between them. Furthermore, the $c_{dis}$ affects the effective shape anisotropy of the particles, which also impacts the nematic phase formation. With increasing the $c_{dis}$ (above 45 mM), the ferrofluid starts to aggregate. After subsequent additions of DBSA to a nematic ferrofluid, we observed that the magneto-optic (MO) response slows down until it completely vanishes.

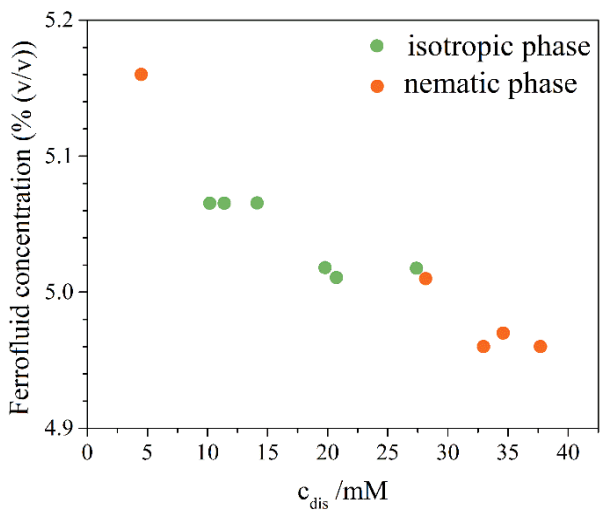

**Fig. 5.** The change of phase upon the addition of DBSA solution. The initial suspension (Sc-middle, cdis = 5.16 %) was at the isotropic-nematic threshold concentration. With a small

addition of the DBSA solution in 1-butanol, the ferrofluid undergoes the phase transition to the isotropic phase, while at high-enough DBSA concentration, i.e. > 25 mM, the ferrofluid reverts to nematic phase again, at ~ 0.2 % (v/v) lower ferrofluid concentration than was the initial one.

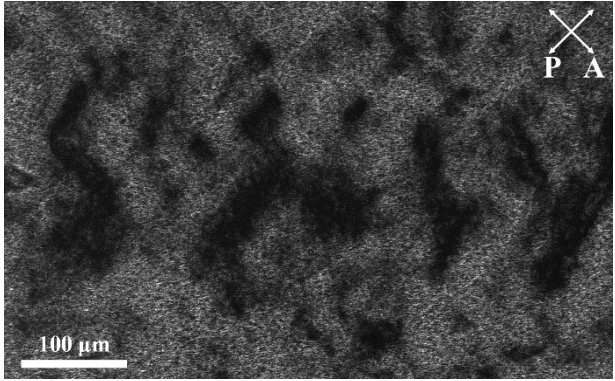

**Fig. 6.** The nematic ferrofluid of the Sc-middle at 5.02 % (v/v) after increased $c_{dis}$.

The MO response of the ordering process starts to decelerate at certain $c_{dis}$ and stops at $c_{dis}$ > 45 mM. At this $c_{dis}$, the ferrofluid cannot be ordered with an external magnetic field of 1.4 mT anymore. The nematic ferrofluids in zero field form magnetic domains as also present in soft ferromagnets [18]. An external magnetic field moves domain walls until the domains are aligned with the external magnetic field. The ordering process is, in our case, observed with POM. When all NPLs are oriented, on average, in the same direction along the capillary (45 ° to polarizer and analyzer), the sample appears bright (Fig. 7). On the other hand, randomly oriented isotropic phase between crossed polarizers appears black (Fig. 7). The effect of a local increase of $c_{dis}$ can also be seen by the naked eye immediately after the addition. A colloidally stable ferrofluid of the BHF NPLs is dark brown with a distinctive shiny look. The addition of DBSA locally makes the ferrofluid brighter and opaque due to different scattering on aggregates than on the colloidally stable NPLs (Fig. 8). After thorough mixing, the ferrofluid becomes shiny again as long as the $c_{dis}$ is low enough that the aggregation does not prevail.

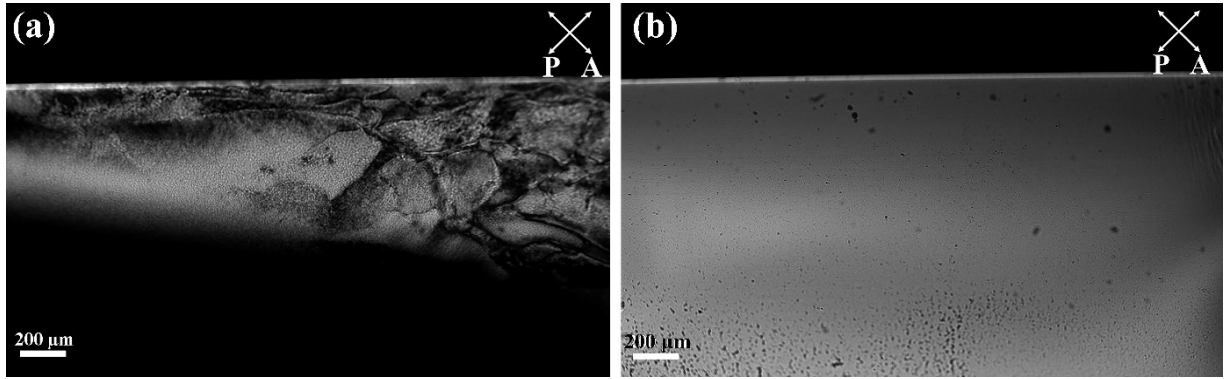

**Fig. 7.** (a) A ferrofluid in zero field, where the top part of the suspension in the capillary is in the nematic phase, (b) MO response (of the same ferrofluid as in (a)) to an external magnetic field applied at 45° to the polarizer.

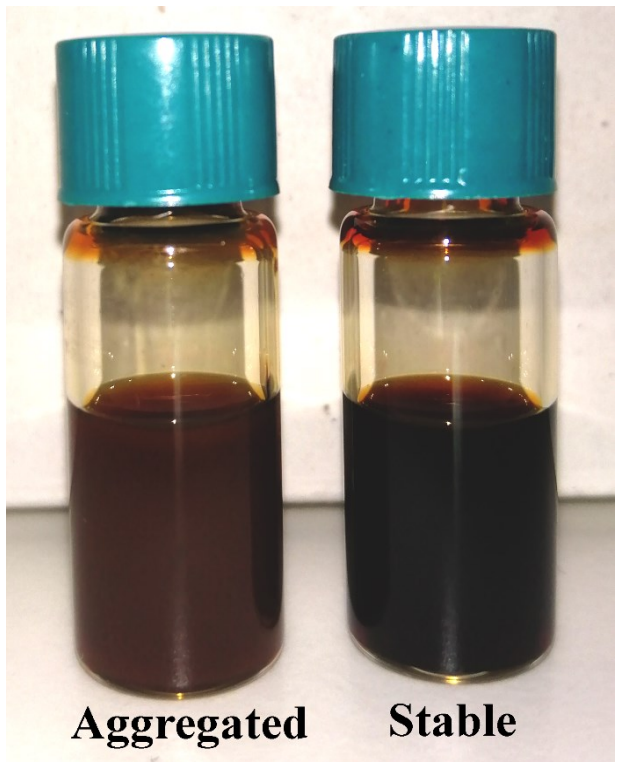

**Fig. 8.** Aggregated (brighter and opaque) ferrofluid on the left and colloidally stable ferrofluid (dark, shiny) on the right side.

### 3.3 Effect of the NPLs saturation magnetization

To evaluate the effect of magnetic interaction strength on the threshold concentration, we compared NPLs with different saturation magnetization values, while the rest of the parameters were kept the same. Since particle size and magnetization are intrinsically correlated, we cannot change the magnetization while keeping the same size distribution of the Sc-substituted NPLs. Thus, we decided to compare the threshold concentration of the Sc-substituted with In-substituted samples. We selected the Sc-bottom and In-bottom ferrofluids because they exhibited comparable size distribution, the same colloidal stabilization (Fig. 3(b), Table 1) and a quite different magnetization (Fig. 9(b), Table 1). The In-bottom has lower saturation magnetization than Sc-bottom, as was initially expected [17]. Both ferrofluids were already in the nematic phase after the centrifugation procedure. The initial concentration of the Sc-bottom was 7.4 % (v/v), while the In-bottom was 8.2 % (v/v). The In-bottom nematic ferrofluid forms similar magnetic domains (Fig. 10) as the Sc-substituted nematic ferrofluids and is therefore in a ferromagnetic phase. Dilution of the In-bottom revealed the threshold concentration at (6.6 ± 0.2) % (v/v), which is ~ 2 % (v/v) higher than is the threshold concentration of the Sc-bottom. This is clear evidence that magnetization has a high impact on the threshold concentration for the isotropic-nematic phase transition [14]. Fig. 11 represents the Sc-bottom and In-bottom threshold concentrations, based on the calculated birefringence (see Supplementary Information S.1).

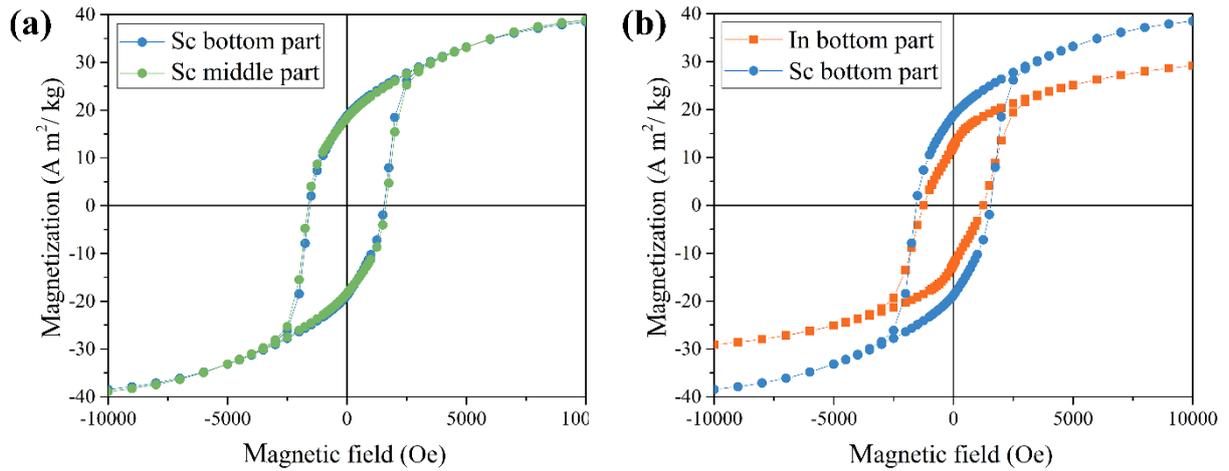

**Fig. 9.** Magnetic hysteresis of (a) Sc-bottom and Sc-middle and (b) Sc-bottom and In-bottom.

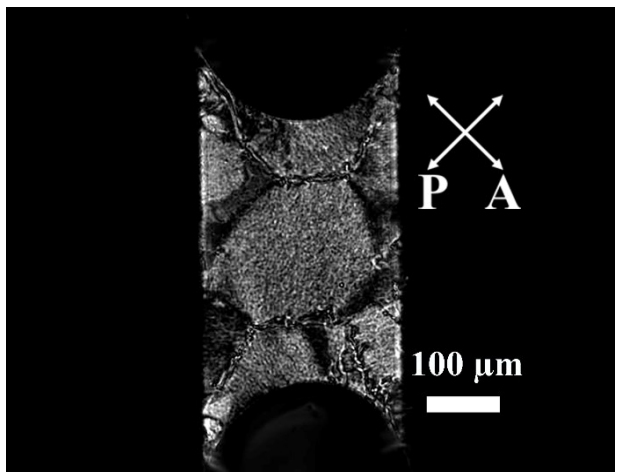

**Fig. 10.** POM image of magnetic domains in the In-bottom in a capillary with dimensions 300 μm × 30 μm in zero field.

Mixing of the Sc- with In-substituted NPLs decreases the sample's overall magnetization compared to the pure Sc-sample. Considering the results from the previous paragraph, we expect that the threshold concentration of the Sc- and In-sample mixture will be in-between, which was also proven as the threshold concentration of the mixture (1:1) was determined to be at (5.2 ± 0.1) % (v/v) (Fig. 11). At the same time, we noticed that the mixed nematic ferrofluids (Sc-bottom and In-bottom samples) were inhomogeneous and revealed some interesting stripe-like structures, not usually observed in the BHF NPLs ferrofluids (Fig. 12(a)).

Under the applied external magnetic field of 1.4 mT, some thread-like long aggregate structures eventually separated. Such behavior made us interested to inspect other mixtures of NPLs that differ more than In-bottom and Sc-bottom samples in the context of interparticle interactions. We took the Sc-bottom and In-middle sample, the latter being in the isotropic phase. The mixing of the two ferrofluids revealed some stripe-like structures already in the absence of an external magnetic field (Fig. 12(b)). The structures quickly separated under the applied magnetic field of 1.4 mT but formed again after 20 minutes in zero field. Such behavior can be explained with the difference in In- and Sc-NPLs magnetization. The Sc-NPLs tend to order into the nematic phase due to stronger ferromagnetic coupling. This results in the phase separation of the ferrofluid mixture, where the Sc-NPLs concentrate in nematic regions with higher orientational order and magnetization, while In-NPLs are present in more disordered areas.

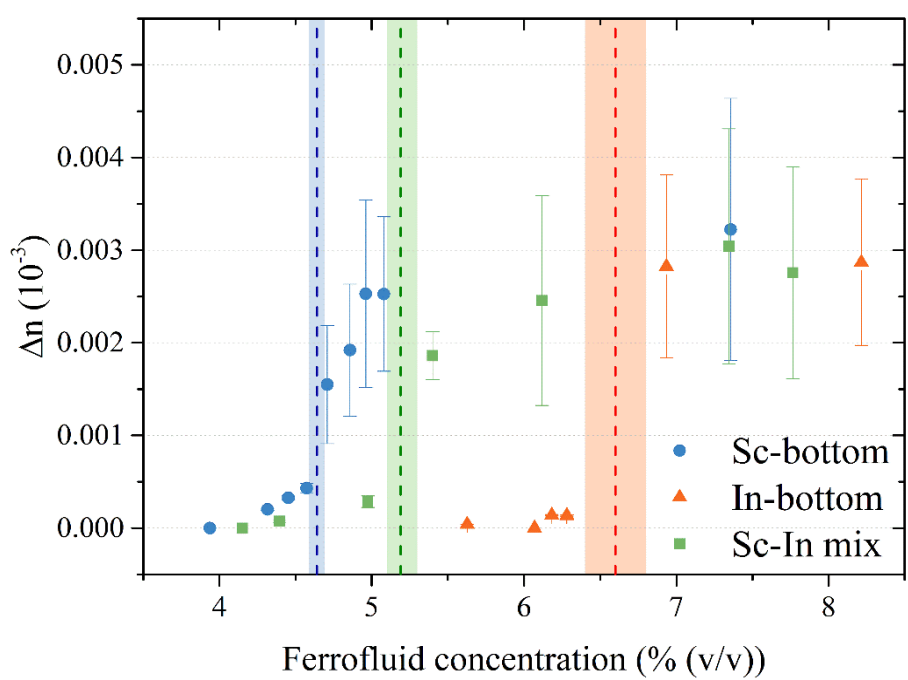

**Fig. 11.** Birefringence with a marked area of the phase transition from isotropic to the nematic phase of the Sc- and In-substituted bottom and Sc-In-bottom mix. The grainier image of the In-bottom nematic phase results in larger error bars. For the measurements of birefringence, the magnetic field was applied along the long axis of the capillary.

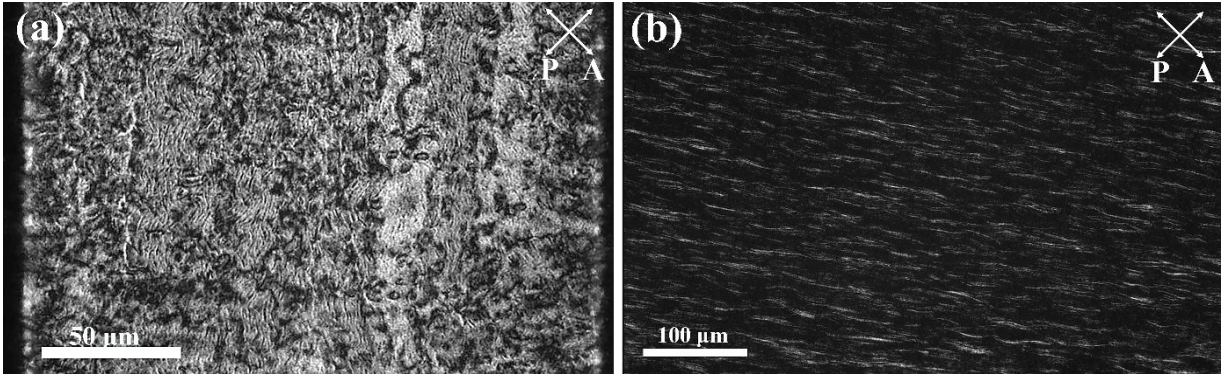

**Fig. 12.** The stripe-like structures in the nematic mixtures of Sc- and In-ferrofluids; (a) In-bottom + Sc-bottom at 7.3 % (v/v) and (b) In-middle + Sc-bottom at 4.7 % (v/v).

**3.4 Polydispersity and interactions between the NPLs**

As shown before, the prevailing interactions between the NPLs in ferrofluids are anisotropic magnetic and electrostatic [16]. While the amount of added surfactant can control the latter, the anisotropic magnetic interactions depend on magnetization determined with the NPL type (Sc- or In-substituted in this case) and particle size. In the studied ferrofluids, NPLs are polydispersed, and thus, interactions are also polydisperse. We will analyze the polydispersity of the interactions between two NPLs with parallel dipole moments separated for a distance $r$ in the direction of the dipoles (see Supplementary Information S.2). In this case, the magnetic attractive interaction energy is the largest, and the total interaction between two different NPLs with magnetic moments $p_{i,j}$ and charges $e_{i,j}$ can be written as

$$U_{tot}(r) = U_{dip}(r) + U_{el}(r) \tag{1}$$

where $U_{dip}(r) = -\frac{\mu_0 p_i p_j}{2\pi r^3}$ and $U_{el}(r) = \frac{e_i e_j}{4\pi\varepsilon\varepsilon_0 r} e^{-\kappa r}$ (see Supplementary Information S.2). Here, $\kappa$ is inverse Debye length. Fig. 13(a) shows total interaction energy averaged over the size distribution of the NPLs (Fig. 3). For comparison also repulsive electrostatic interaction energy is shown. Fig. 13(b), (c) and (d) presents the total interaction energy for all combinations of the NPLs sizes in histograms (Fig. 3), where black lines is the average interaction energy. Because the size distribution is relatively large (Fig. 3), the interaction between different size classes varies in amplitude and changes from repulsive, for smaller, to attractive, for larger NPLs (Fig. 13(b), (c), (d)).

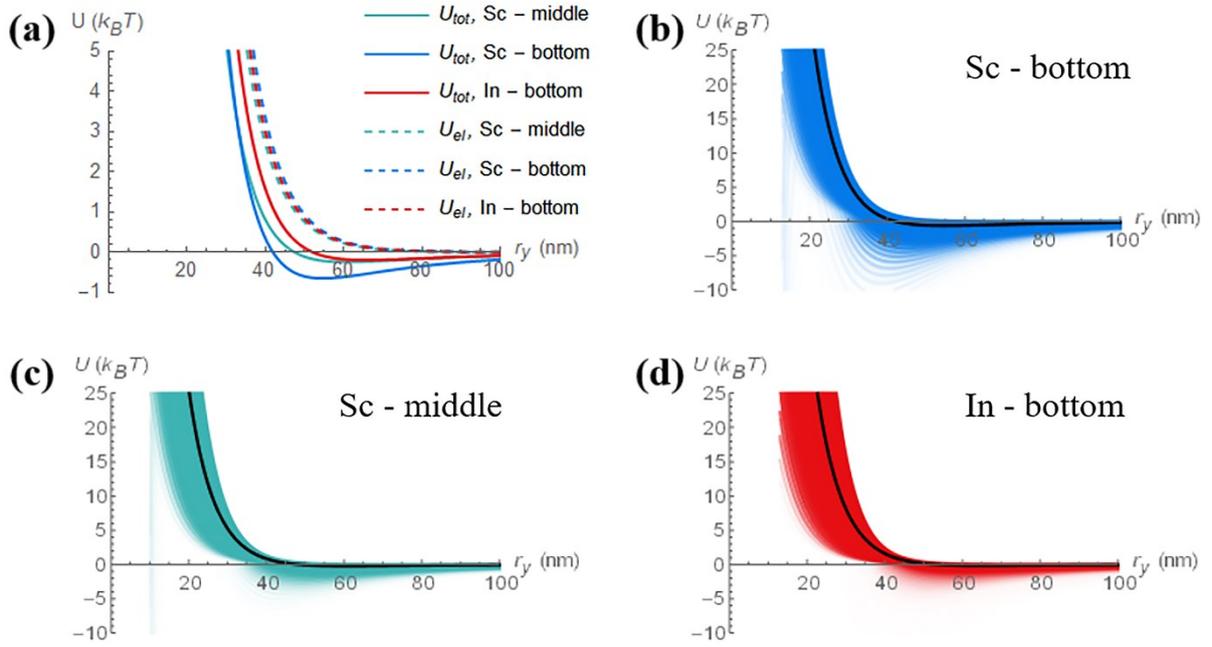

**Fig. 13.** (a) Calculated interaction energy between a pair of disks with Eq. (1) averaged over the size of the NPLs for parallel dipole moments (see text) for the three ferrofluids as marked. The polydispersity of the interaction: (b), (c), (d) shows total interaction energy for all combinations of the NPLs sizes in histograms (Fig. 3). The opacity of the lines is proportional to the probability for a given pair. Black lines show average interaction energy. The Debye screening length is, in all cases, 9.5 nm.

## 4. Discussion

Shape anisotropy of the particles plays an essential role in the liquid crystal behavior of suspensions. Onsager explained the nematic ordering of the suspensions with anisotropic particles [12]. At higher concentrations packing entropy overrules orientational entropy, making favorable conditions for the particles to form a nematic phase. He also demonstrated that the nematic ordering could be induced by the anisotropic shape only, which was also supported with computer simulations [19,20]. Platelet systems show various phases, such as nematic, lamellar, and columnar, e.g. [8,21,22]. Platelet suspensions undergo the transition to the nematic phase at a specific threshold concentration. Computer simulations have shown that for thin nonmagnetic platelets without surface charge, the transition to the nematic phase happens at an effective volume fraction (also called dimensionless transition density) $\phi_{NI,eff} \approx 3-4$, depending on the shape of a platelet [23]. The effective volume fraction is calculated as $\phi_{NI,eff} = \frac{N}{V}D^3$, where $D$ is the equivalent diameter of the platelet and is related to the excluded volume. For platelets of finite thickness, the threshold concentration is considered to be largely independent of the aspect ratio ($h/D$). For monodispersed nonmagnetic platelets, when $h/D > 1/7$, the system exhibits a direct transition from the isotropic to the columnar phase [19]. If the platelets have a surface charge, they are effectively larger for about Debye screening length in each direction, so the volume fraction of the platelets is effectively larger and the aspect ratio ($h/D$) effectively smaller, which affects their phase behavior [8,24]. Intermediate range electrostatic repulsion ($\kappa D \sim 30\text{-}100$) destabilizes the columnar phase in favor of nematic order. In contrast, for longer-range repulsions ($\kappa D < 7$), a destabilization of the nematic phase in favor of lamellar order is found [22]. The existence of different (i.e., nematic, lamellar, and columnar) liquid crystal phases is therefore controlled by the strength and range of the double-layer repulsion [22].

Table 2 shows the experimentally determined threshold concentrations $\phi_{NI}$, the calculated effective volume fractions $\phi_{NI,eff}$, and the rescaled volume fractions $\phi_{NI,resc}$ and $\phi_{NI,eff,resc}$ considering the Debye screening length. The $\lambda_D$ was determined from the $c_{dis}$ to be 9.5 nm [16], which gives $\kappa D$ around 5 (Table 2). The aspect ratio of the BHF NPLs depends on the diameter of the NPLs because the NPLs thickness does not vary significantly [16,25]. While the effective volume fractions are very small, 1.1 for Sc-ferrofluids and 1.6 for In-bottom, the rescaled ones (2.4–3.4) are much more similar to the threshold values in a nonmagnetic system [8,23,26] (Table 2). Higher magnetic interaction strength lowers the threshold concentrations.

**Table 2.** The experimentally determined threshold concentrations and calculated effective aspect ratio, magnetic moment, surface charge, effective and rescaled volume fractions, where $D_{eff} = D + 2\lambda_D$, $h_{eff} = h + 2\lambda_D$, and $V_{eff} = \frac{\pi}{4} D_{eff}^2 h_{eff}$.

|  | Sc-middle | Sc-bottom | In-bottom |
|---|---|---|---|
| $\phi_{NI}$ | 0.052 | 0.046 | 0.066 |
| $\langle h/D \rangle$ | 0.10 | 0.08 | 0.09 |
| $\langle p_m \rangle$ ($10^{-18}$ A m$^2$) | 1.37 | 1.96 | 1.35 |
| $\langle e \rangle$ (e$_0$) | 43 | 52 | 49 |
| $\phi_{NI,eff} = \frac{N}{V} \langle D^3 \rangle$ | 1.1 | 1.1 | 1.6 |
| $\phi_{NI,resc} = \frac{N}{V} \langle V_{eff} \rangle$ | 0.20 | 0.16 | 0.23 |
| $\phi_{NI,eff,resc} = \frac{N}{V} \langle D_{eff}^3 \rangle$ | 2.6 | 2.4 | 3.4 |
| $\langle \kappa D \rangle$ | 4.7 | 5.7 | 5.5 |

The dipolar magnetic interaction is strongly anisotropic and (on average) attractive. On its own, it directs the assembly of disks into columns, and without sufficient electrostatic repulsion, causes aggregation. A ferromagnetic nematic ordering is a complicated many-body problem. However, some features can be understood from the pair interaction, a combination of magnetic and screened electrostatic interaction. It can be purely repulsive, or it can be repulsive at short range but attractive at larger distances, which results in a minimum (Fig. 13). This minimum, when deep enough, causes the formation of columns [27]. The SANS and SAXS measurements of this system show that, while the minimum in the interaction causes positional and orientational correlations, it does not result in the formation of longer columns in the considered systems [14,28]. There are two reasons for that. The first and main reason is the centrifugation during the preparation of the samples in which aggregates and larger columns are discarded. The second reason is the large polydispersity (40 %) in NPLs' diameter, known to disfavor columnar order [21] to avoid the effect of polydispersity in favor of the lamellar phase. The polydispersity in diameter is also reflected in the polydispersity of the interaction (Fig. 13(b)-(d)), which is most likely the underlying cause of the phase separation observed in the mixtures of In- and Sc-samples. Platelets with larger magnetic moments (Sc-sample) attract each other and form elongated islands of the more ordered nematic phase. At the same time, those for which repulsive electrostatic interaction prevails (In-sample) remain in a less ordered surrounding phase.

The electrostatic interaction is determined by two parameters, the surface charge and the Debye screening length. The first determines the magnitude and the second the range of the interaction. Both have similar effects on the phase transition. Comparison between In-bottom and Sc-middle, in which platelets have similar average dipole moments but are of different diameters and, consequently, of different average charges, shows that smaller charge causes the transition to nematic phase at much smaller concentrations. Effectively, the decrease of charge allows NPLs to get closer to form better-oriented assemblies/longer columns (Fig. 13(a)). On the other hand, the addition of the DBSA, which reduces the Debye length, only slightly reduces the threshold concentration. In this case, as shown in Fig. S3, not only can a pair of NPLs get closer at smaller Debye lengths, but they can also get trapped in a minimum of the interaction. This process leads to reversible destabilization of the ferrofluid.

## 5. Conclusions

We studied how the isotropic-nematic phase transition of ferrofluids made of barium hexaferrite nanoplatelets is affected by the size distribution, magnetization of the nanoplatelets and ionic strength. These are all parameters that also affect the colloidal stability of the ferrofluid. The results show that the transition from the isotropic to nematic phase is dominated by the interplay between the attractive dipolar magnetic and repulsive screened electrostatic interaction. The increase of the magnitude of magnetic interaction or decrease in magnitude or/and range of the electrostatic interaction cause that the interaction changes from repulsive to attractive at shorter distances and shifts the transition to the nematic phase to lower concentration.

Among the studied parameters, magnetization had the most significant impact on the threshold concentration. At the same time, the results show that the combination of all parameters has to be in a delicate balance to provide attractive interaction at the intermediate range, which favors nematic ordering, and strong enough short-ranged repulsive interaction to prevent aggregation. These findings will help to design and optimize new materials exhibiting the ferromagnetic liquid phase.

Future investigations of this system will include computer simulations that could verify the existence of phases that we have not detected in the currently studied system, e.g. columnar and lamellar, present in some nonmagnetic systems.

**Supplementary Information:** S.1 Description of the calculations for birefringence and a figure, representing an example of intensity and birefringence for Sc-substituted BHF NPLs. S.2 Description of the calculations of interactions between the NPLs and a figure representing the dependence of average interaction of parallel platelets on the Debye length for Sc-middle ferrofluid.

**Declaration of Competing Interests**

The authors declare that they have no competing interests.

**Acknowledgements**

This work was financially supported by the Slovenian Research Agency (P1-0192, P2-0089, J1-2459, PR-08973 and PR-08415). We thank the Center of excellence in nanoscience and nanotechnology (CENN) Nanocenter for the use of TEM Jeol 2100 and Lakeshore 7404 vibrating-sample magnetometer.

**Funding**

This work was supported by the Slovenian Research Agency (P1-0192, P2-0089, J1-2459, PR-08973 and PR-08415).

**Author's contributions**

The study was conceptualized by AM and DL. PHB wrote the paper, synthesized the nanoplatelets, prepared the ferrofluids, performed magnetic and conductivity measurements, took the transmission electron microscopy images and determined the size distributions. ŽG set the setup for the polarizing optical microscopy observation and performed birefringence measurements and calculations. NO and ŽG designed and constructed the custom-made 3-axis magnetic field manipulation system. NS and AM guided the interpretation of the data to obtain birefringence and the final results in Fig. 11. AM calculated the results presented in 3.4. AM/DL/NS carefully revised the manuscript and importantly improved the final manuscript. All authors have read and approved the final manuscript.